\documentclass[conference]{IEEEtran}
\IEEEoverridecommandlockouts
\usepackage{cite}
\usepackage{amsmath,amssymb,amsfonts}
\usepackage{algorithmic}
\usepackage{graphicx}
\usepackage{textcomp}
\usepackage{xcolor}
\usepackage{subcaption}
\def\BibTeX{{\rm B\kern-.05em{\sc i\kern-.025em b}\kern-.08em
    T\kern-.1667em\lower.7ex\hbox{E}\kern-.125emX}}

\newcommand{\m}{{\bf m}} 
\newcommand{\x}{{\bf x}} 
\newcommand{\y}{{\bf y}} 
\newcommand{\X}{{\bf X}} 
\newcommand{\Y}{{\bf Y}} 
\renewcommand{\H}{{\bf H}} 
\newcommand{\h}{{\bf h}} 
\newcommand{\I}{{\bf I}} 
\newcommand{\C}{{\bf C}} 
\newcommand{\Z}{{\bf Z}} 
\newcommand{\W}{{\bf W}} 
\newcommand{\bias}{{\bf b}} 
\newcommand{\E}{\mathbb{E}} 
\newcommand{\KL}{\mathrm{KL}} 

\DeclareMathOperator*{\argmax}{arg\,max}

\begin{document}

\title{Learning to Modulate for Non-coherent MIMO}

\author{\IEEEauthorblockN{Ye Wang}
\IEEEauthorblockA{\textit{Mitsubishi Electric Research Laboratories}\\
Cambridge, MA, USA \\
yewang@merl.com}
\and
\IEEEauthorblockN{Toshiaki Koike-Akino}
\IEEEauthorblockA{\textit{Mitsubishi Electric Research Laboratories}\\
Cambridge, MA, USA \\
koike@merl.com}
}

\maketitle

\begin{abstract}
The deep learning trend has recently impacted a variety of fields, including communication systems, where various approaches have explored the application of neural networks in place of traditional designs. Neural networks flexibly allow for data/simulation-driven optimization, but are often employed as black boxes detached from direct application of domain knowledge. Our work considers learning-based approaches addressing modulation and signal detection design for the non-coherent MIMO channel. We demonstrate that simulation-driven optimization can be performed while entirely avoiding neural networks, yet still perform comparably. Additionally, we show the feasibility of MIMO communications over extremely short coherence windows (i.e., channel coefficient stability period), with as few as two time slots.
\end{abstract}


\section{Introduction}

The application of machine learning techniques to communication systems has recently received increased attention~\cite{o2016learning, o2017introduction, o2017physical, erpek2018learning, kim2018communication, kim2018deepcode, karanov2018end, dorner2018deep, aoudia2018end, ye2018power, raj2018backpropagating, o2018physical}.
Common to these approaches is the data/simulation-driven optimization of neural networks (NN) to serve as various communication system components, instead of traditional approaches that are systematically driven by models and theory.
The promise of such approaches is that learning could potentially overcome situations where limited models are inaccurate and complex theory is intractable.
This can be viewed as part of a larger ``deep learning'' trend, where the enthusiastic application of modern machine learning methods, revolving around deep neural networks, have widely impacted a variety of fields~\cite{lecun2015deep}.

We consider an end-to-end, learning-based approach to optimize the modulation and signal detection for non-coherent, multiple-input, multiple-output (MIMO) systems, i.e., communication with multiple transmit and receive antennas, where the channel coefficients are unknown.
The {\em end-to-end} aspect refers to the joint optimization of both the signal constellation and decoder as it would interact with a simulated MIMO channel to transmit and receive messages.
As noted in the literature~\cite{o2016learning, o2017introduction}, this general concept is analogous to training an autoencoder, but with a noisy channel inserted between the encoder and decoder, which has led several works~\cite{o2016learning, o2017introduction, o2017physical, erpek2018learning, kim2018deepcode, karanov2018end, dorner2018deep, aoudia2018end, raj2018backpropagating, o2018physical} to use deep neural networks to realize both the encoder and decoder mappings.
Related work~\cite{o2017physical} and~\cite{erpek2018learning} also consider the MIMO channel, although with channel state information (CSI) available, and the latter also examines a multi-user interference channel.

One aim of our paper is to reconsider the benefits of employing neural networks and demonstrate an effective learning-based approach that eschews them altogether.
Mapping from a finite message space to channel symbols does not require a neural network encoder, since a lookup table storing the signal constellation is sufficient.
Non-coherent MIMO decoding theory~\cite{hochwald2000unitary} guides us to a simplified decoder architecture that avoids employing neural networks, while still retaining the ability to perform simulation-driven optimization.
We evaluate and compare this network-less approach versus employing a neural network decoder, and find that they perform comparably, although additional hyperparameter tuning for both could potentially further improve performance and stability.

We also use our learning-based approach to demonstrate that non-coherent MIMO communication is feasible even at extremely short coherence windows, i.e., with the channel coefficients stable for as few as two time slots.
Unlike various conventional approaches~\cite{hochwald2000unitary, hochwald2000systematic, liang2002unitary, koike2010high} to MIMO modulation design, which require limitations on time slots versus antennas, the learning-based approach is not so limited by analytical design constraints.
Relaxing these constraints is also supported by the recent extension by~\cite{yang2013capacity} of MIMO capacity theory~\cite{telatar1999capacity, marzetta1999capacity}, which shows that the conventional unitary, isotropically distributed inputs are no longer capacity achieving when antennas exceed time slots.

\subsection{Notation}

We use uppercase/lowercase bold letters, e.g., $\X$ and $\m$, to denote matrices/vectors.
A circularly-symmetric Gaussian distribution with zero mean and $\sigma^2$ variance is denoted by $\mathcal{CN}(0, \sigma^2)$.
We write $\X^\dagger$ to denote the conjugate transpose of $\X$, and $\I_m$ to denote the $m \times m$ identity matrix.

\section{Modulation Optimization for MIMO Systems}

\subsection{Non-Coherent MIMO Channel}

We consider transmission over a MIMO channel with $m$ transmitter antennas and $n$ receiver antennas.
When transmitting a message using $L$ channel symbols, the received signal $\Y$ is an $n \times L$ complex matrix given by
\begin{align*}
\Y = \H \X + \Z,
\end{align*}
where $\X$ is an $m \times L$ complex matrix representing the transmitted signal, $\H$ is the $n \times m$ complex, random channel matrix, and $\Z$ is an $n \times L$ complex matrix representing Gaussian noise.
We focus on the {\em non-coherent} case where the random channel matrix $\H$ is unknown (i.e., no CSI), but fixed over the $L$ channel uses.
The elements of channel $\H$ are i.i.d. $\mathcal{CN}(0, 1/m)$ and are independent of the noise $\Z$, which is i.i.d. $\mathcal{CN}(0, \sigma^2)$.
We constrain the transmission to have average power $\E[\| \X \|^2/(mL)] = 1$, such that the signal-to-noise (SNR) ratio is given by $1 / \sigma^2$.

\subsection{Encoder Parameterization}

The encoder maps a $k$-bit message to an $L$ symbol transmission across $m$ antennas.
This encoder mapping, $f: \{0, 1\}^k \rightarrow \mathbb{C}^{m \times L}$, can be generally parameterized by a simple lookup table specified by a codebook matrix $\C \in \mathbb{C}^{2^k \times mL}$.
For power efficiency, the mean row of $\C$ is subtracted from each row of $\C$ to produce the centered codebook matrix $\overline{\C}$.
Then, the average power constraint is enforced by scaling $\overline{\C}$ to produce centered and normalized code matrix $\widetilde{\C} := \overline{\C} \sqrt{ (2^k mL) / \|\overline{\C}\|^2}$.
To encode a message $\m \in \{0, 1\}^k$, the encoder mapping selects the row in $\widetilde{\C}$ indexed by the integer value of $\m$, and reshapes it to an $m \times L$ matrix to form the transmitted signal $\X_\m := f_\C(\m) \in \mathbb{C}^{m \times L}$.
Essentially, this entire procedure is just to allow the signal constellation $\{\X_\m\}_{\m \in \{0,1\}^k}$, which is constrained in its average power, to be parameterized by the unconstrained variable $\C$.

\subsection{Decoder Realizations} \label{sec:decoders}

We consider two parametric, soft-output decoders that approximate the unnormalized, log-likelihoods for each possible message, and thus output a real vector of length $2^k$.
For both decoders, the softmax operation is applied to the output vector (by exponentiating each element and then scaling to normalize the sum to one) to produce a stochastic vector, denoted by $P^\theta_{\m | \Y}$, that approximates the posterior distribution $P_{\m | \Y}$.
Note that applying the softmax operation to the vector of unnormalized, log-likelihoods $\{ \log \alpha P_{\Y|\m}(\Y|\m) \}_{\m \in \{0,1\}^k}$, for some constant $\alpha > 0$, would yield the corresponding posterior distribution $\{ P_{\m|\Y}(\m|\Y) \}_{\m \in \{0,1\}^k}$.

\subsubsection{Pseudo-ML (pML) Decoder} \label{sec:pML-decoder}

If the codewords are orthonormal, that is, $\X_\m \X_\m^\dagger = L \I_m$ for all messages $\m \in \{0, 1\}^k$, then the ML decoding rule is shown in~\cite{hochwald2000unitary} to be
\begin{align} \label{eqn:unitaryML-decoding}
\argmax_{\m \in \{0, 1\}^k} \|\Y \X_\m^\dagger\|^2,
\end{align}
since the terms $\|\Y \X_\m^\dagger\|^2$ are proportional to $\log \alpha P(\Y | \m)$, for some $\alpha > 0$ that is constant with respect to $\m$.
This decoder immediately inspires a soft-output decoder that simply scales the objective in~\eqref{eqn:unitaryML-decoding} with a parameter $\theta \geq 0$ to output
\begin{align} \label{eqn:softML}
\{ \theta \|\Y \X_\m^\dagger\|^2 \}_{\m \in \{0,1\}^k}.
\end{align}
The parameter $\theta$ both accounts for the fact that $\|\Y \X_\m^\dagger\|^2$ is only proportional to $\log \alpha P(\Y | \m)$, and allows the confidence of the decoder to be tuned, which is particularly important since it will be employed while enforcing the orthonormal constraint (i.e., $\X_\m \X_\m^\dagger = L \I_m$) in only a soft manner.
Hence, we call this the {\em pseudo}-ML (pML) decoder.
Smaller/larger $\theta$ indicates lower/higher confidence, as the corresponding posterior estimate $P^\theta_{\m | \Y}$ (produced by applying the softmax operation) approaches uniform as $\theta \rightarrow 0$ and certainty as $\theta \rightarrow \infty$.

\subsubsection{Neural Network (NN) Decoder}

Alternatively, a soft-output decoder can be realized with a neural network, which serves as a parametric approximation for the mapping
\begin{align} \label{eqn:pureNN-decoding}
g_\theta : \mathbb{C}^{n \times L} \rightarrow \mathbb{R}^{2^k},
\end{align}
where $\theta$ denotes the parameters specifying the weights of the neural network layers. 
The network is applied to the received signal to yield an approximation of the log-likelihoods, to which the softmax operation is applied to produce the corresponding posterior estimate $P^\theta_{\m | \Y} := \mathrm{SoftMax}(g_\theta(\Y))$.

The specific network architectures used in our experiments are detailed alongside discussion of the results in Section~\ref{sec:archs}.
In order to handle a complex-valued matrix as input, $\Y$ is simply decomposed into its real and imaginary components and vectorized, i.e., $\Y$ is represented as a real vector of length $2nL$.

\subsection{Optimization Objective} \label{sec:obj}

The main optimization objective is to minimize the cross-entropy loss with respect to the encoder and decoder parameters, as given by
\begin{align} \label{eqn:NN-objective}
\min_{\C, \theta} \E[ - \log P^\theta_{\m | \Y}(\m | \Y) ],
\end{align}
where $P^\theta_{\m | \Y}$ is produced by applying the softmax operation to the log-likelihoods produced by either decoder given by~\eqref{eqn:softML} or~\eqref{eqn:pureNN-decoding}, as described in Section~\ref{sec:decoders}.
Since the cross-entropy loss can be written as
\begin{align*}
\E[ - \log P^\theta_{\m | \Y}(\m | \Y) ] = H(\m | \Y) + \KL(P_{\m | \Y} \| P^\theta_{\m | \Y}),
\end{align*}
the ideal optimization of the decoder should cause the estimated posterior $P^\theta_{\m | \Y}$ to converge toward the true posterior $P_{\m | \Y}$, and the overall optimization is equivalent to maximizing the mutual information $I(\m; \Y) = H(\m) - H(\m | \Y)$, with respect to the signal constellation, since $H(\m) = k$ is constant.

As mentioned earlier, the pML decoder given by~\eqref{eqn:softML} is formulated assuming orthonormal codewords that satisfy $\X_\m \X_\m^\dagger = L\I_m$ for all $\m \in \{0,1\}^k$.
We enforce orthonormality as a soft constraint by introducing an additional {\em orthonormal-loss} term given by
\begin{align*}
\ell(\C) := \frac{1}{2^k m^2}\sum_{\m \in \{0, 1\}^k} \| \X_\m \X_\m^\dagger/L - \I_m \|^2.
\end{align*}
The optimization objective that we use for the pML decoder is formed by combining this orthonormal loss with the primary cross-entropy loss as follows
\begin{align} \label{eqn:ML-objective}
\min_{\C, \theta} \E[ - \log P^\theta_{\m | \Y}(\m | \Y) ] \big( 1 + \lambda \ell(\C) \big),
\end{align}
where $\lambda > 0$ is a weighting parameter to control the impact of the orthonormal loss term.
Note that rather simply adding on the orthonormal loss term, i.e., using an objective of the form $\E[ - \log P^\theta_{\m | \Y}(\m | \Y) ] + \lambda \ell(\C)$, the loss terms have been multiplicatively combined in~\eqref{eqn:ML-objective}.
We found from experimentation that this improved the reliability of convergence, possibly since these loss terms might decay at very different rates making it difficult to tune the hyperparameter $\lambda$ in an additive combination.



\begin{figure*}[ht!]
\centering
\begin{subfigure}{0.49\textwidth}
    \includegraphics[width=\textwidth]{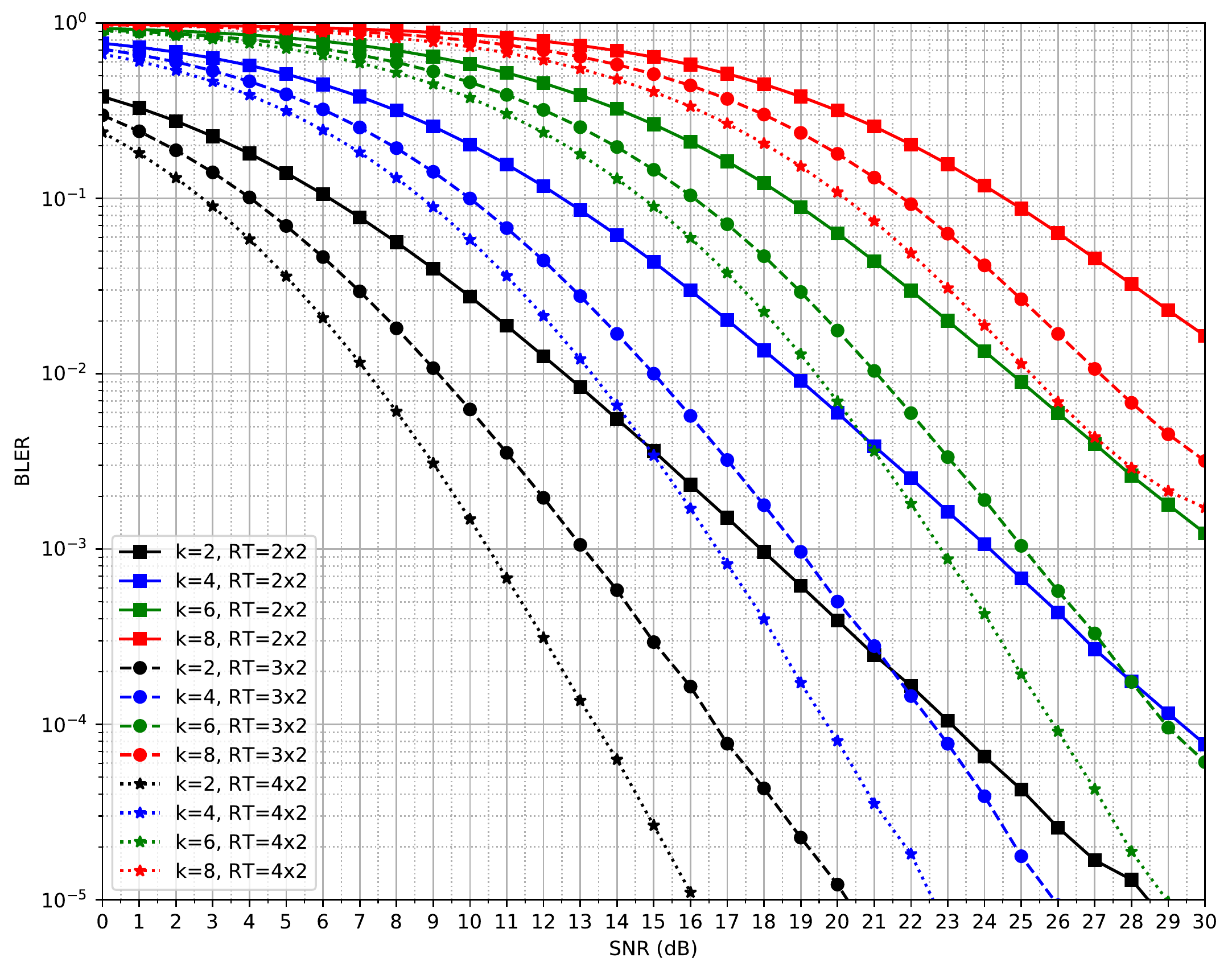}
    \caption{Neural Network (NN) Decoder} 
\end{subfigure}
\begin{subfigure}{0.49\textwidth}
    \includegraphics[width=\textwidth]{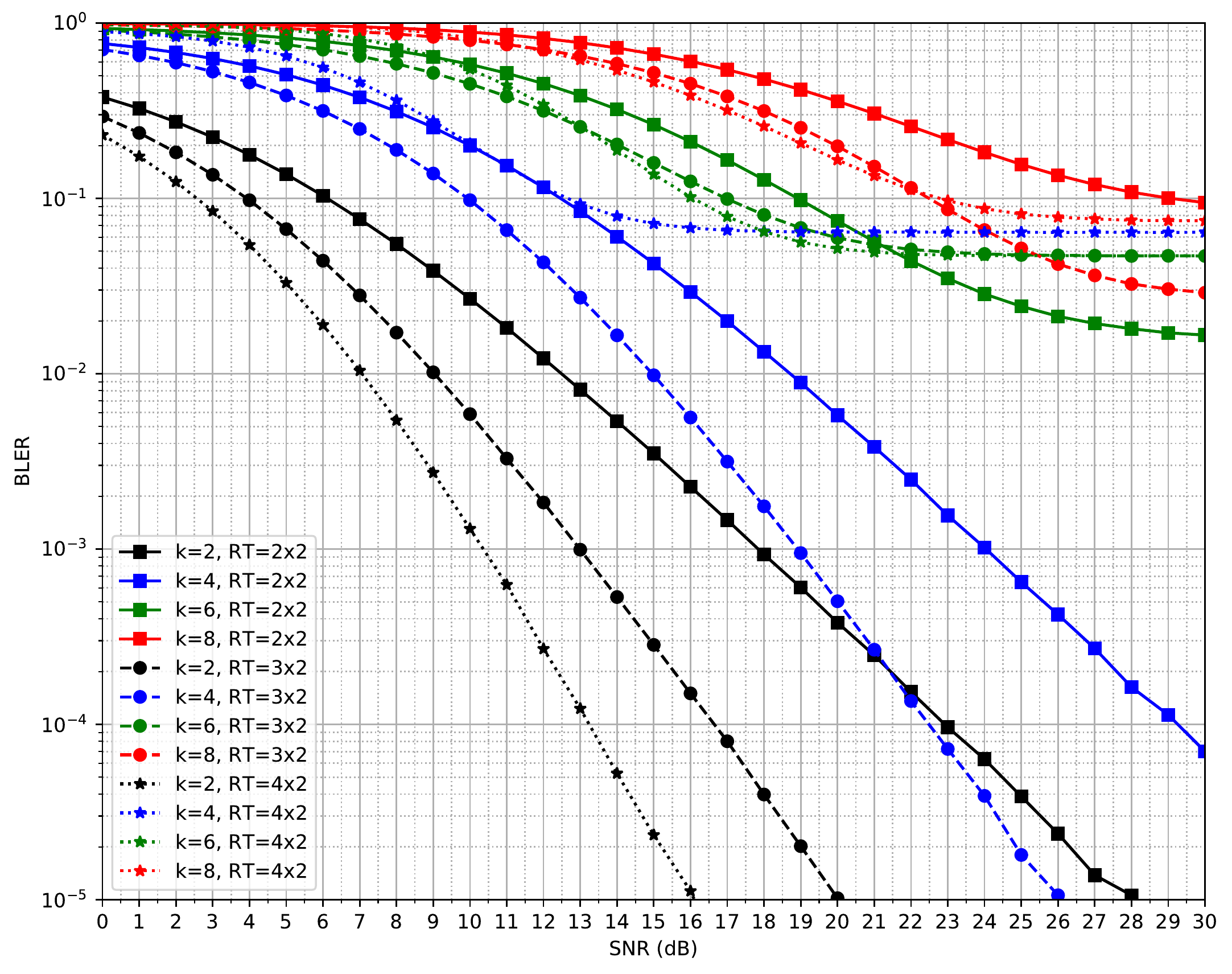}
    \caption{Pseudo-ML (pML) Decoder} 
\end{subfigure}
\caption{BLER performance comparison for $L=2$, with (a) Neural Network Decoder, (b) Pseudo-ML Decoder.}
\label{fig:BLER-L2}
\end{figure*}

\begin{figure*}[ht!]
\centering
\begin{subfigure}{0.49\textwidth}
    \includegraphics[width=\textwidth]{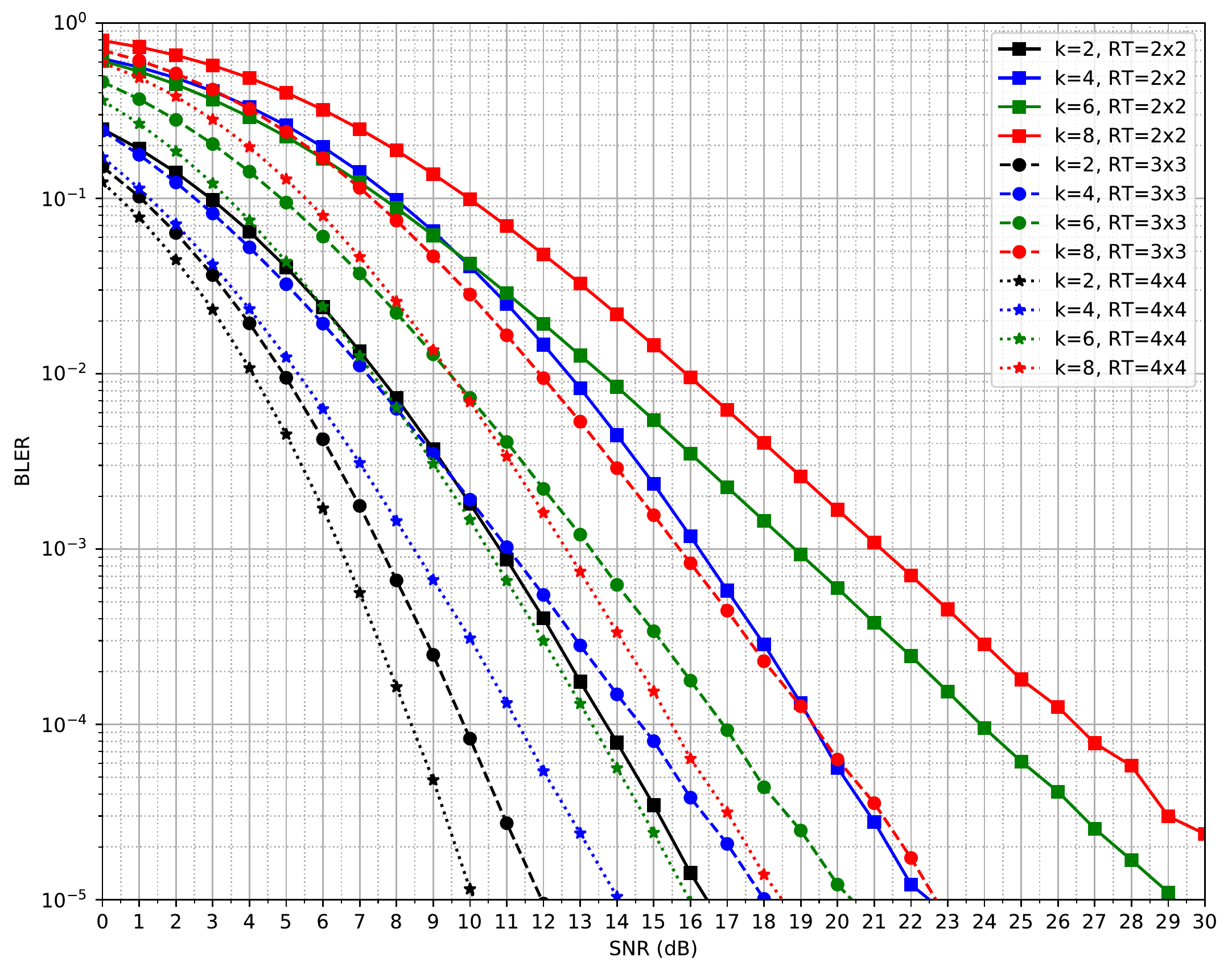}
    \caption{Neural Network (NN) Decoder} 
\end{subfigure}
\begin{subfigure}{0.49\textwidth}
    \includegraphics[width=\textwidth]{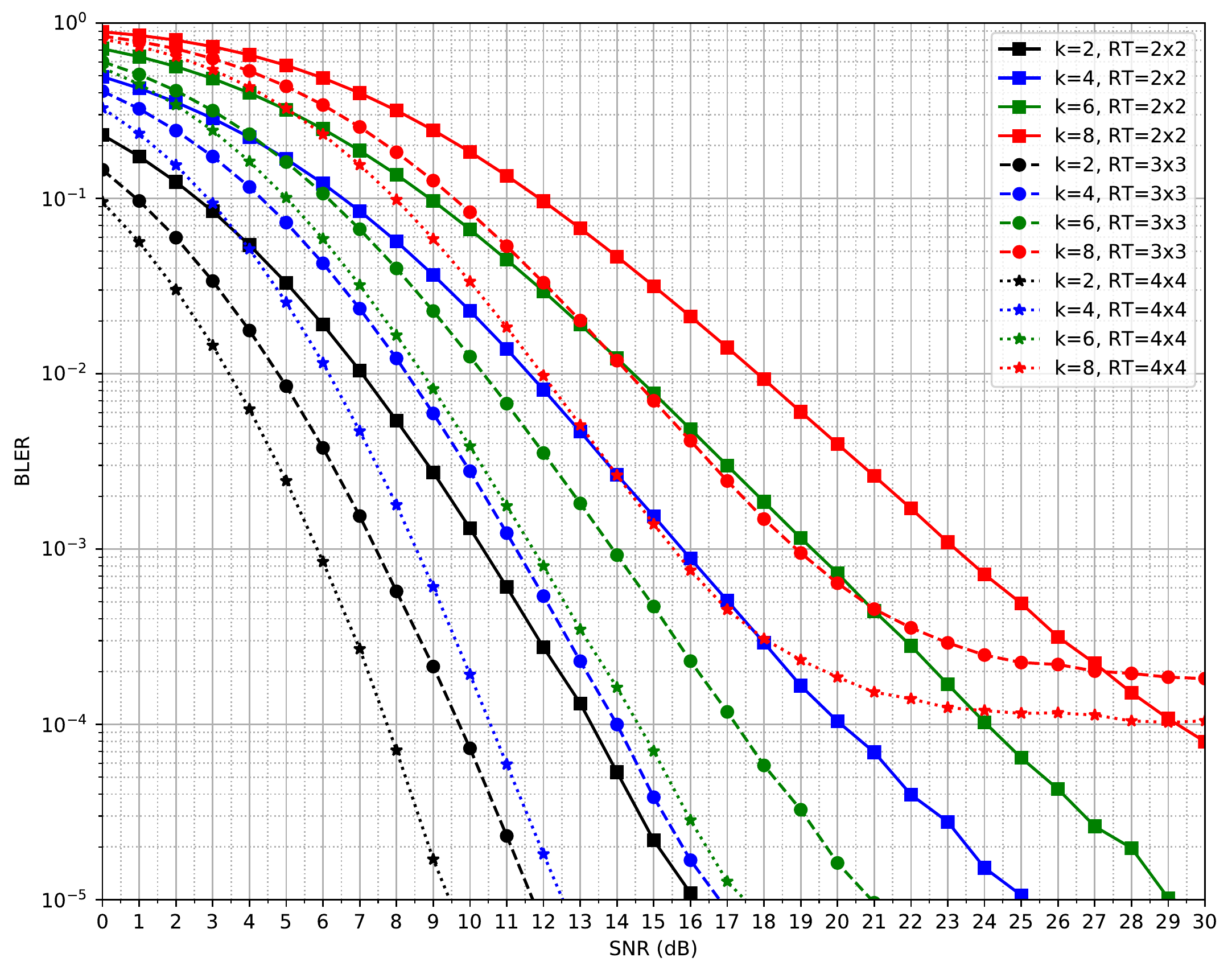}
    \caption{Pseudo-ML (pML) Decoder} 
\end{subfigure}
\caption{BLER performance comparison for $L=4$, with (a) Neural Network Decoder, (b) Pseudo-ML Decoder.}
\label{fig:BLER-L4}
\end{figure*}

\section{Experiments and Results}

Our experiments evaluate communicating $k \in \{2, 4, 6, 8\}$ bits over $L \in \{ 2, 4 \}$ channel uses.
For $L = 2$ time slots, we vary the number of receiver antennas $n \in \{2, 3, 4\}$, while keeping the number of transmit antennas fixed at $m = 2$, since theory~\cite{telatar1999capacity, marzetta1999capacity} teaches that unilaterally increasing transmit antennas $m > L$ does not increase capacity.
We did also test increasing $m > L$ and found that it resulted in performance nearly identical to $m = L$.
For $L = 4$ time slots, we vary both the number of transmit and receive antennas $(m, n) \in \{ (2,2), (3,3), (4,4) \}$.
For each operating point (combination of parameters $k, L, m, n$), we evaluated both the pML and NN decoders, by optimizing each across a variety of hyperparameters, and selecting the best performing codes.
Further details about the network architectures and training procedures are given in Sections~\ref{sec:archs} and~\ref{sec:hyper}.

\begin{figure*}[ht!]
\centering
\begin{subfigure}[t]{0.48\textwidth}
    \includegraphics[width=\textwidth]{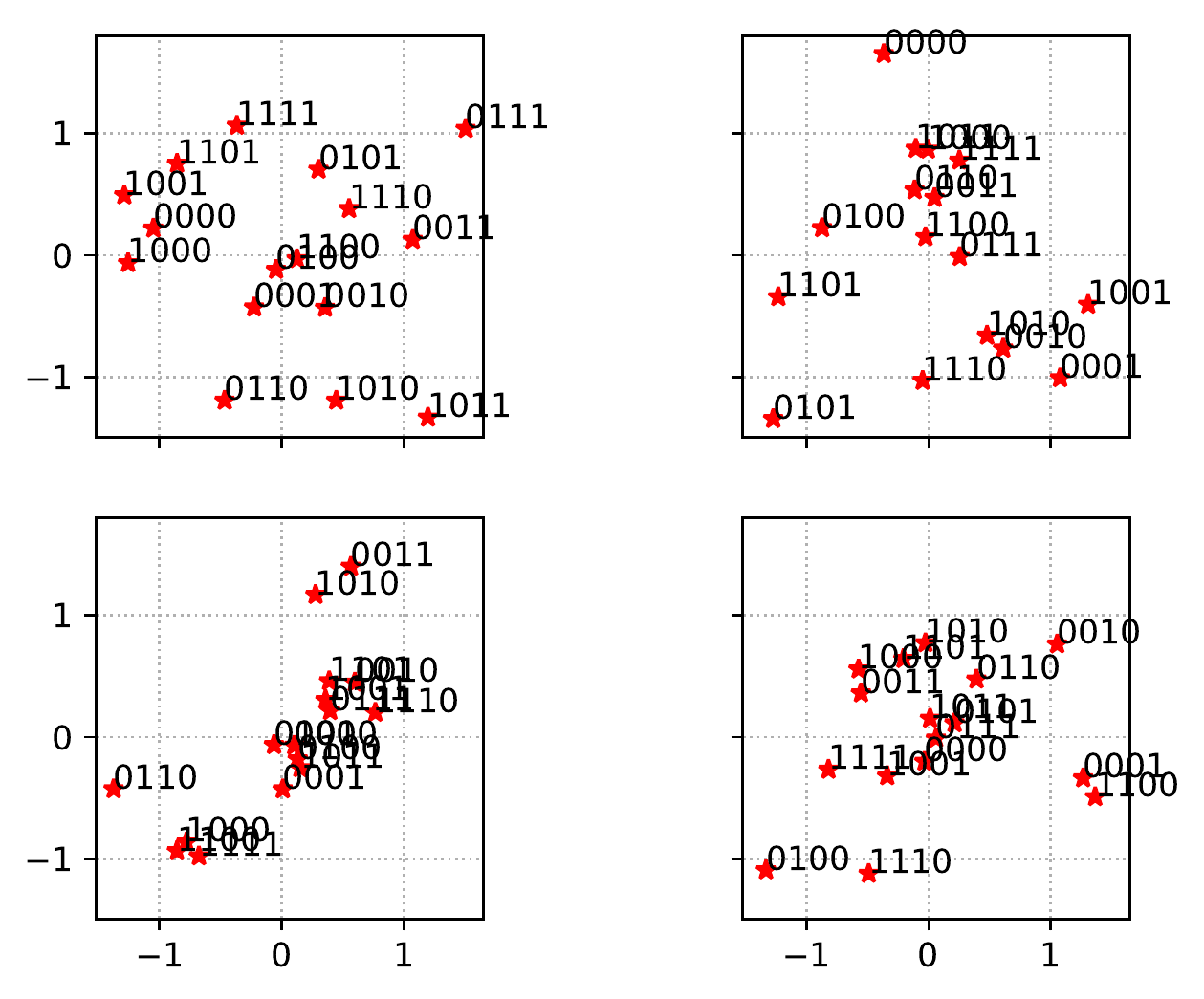}
    \caption{Learned with NN decoder for $k=4$, $L=2$, $(m,n)=(2,3)$}
\end{subfigure}
\begin{subfigure}[t]{0.5\textwidth}
    \includegraphics[width=\textwidth]{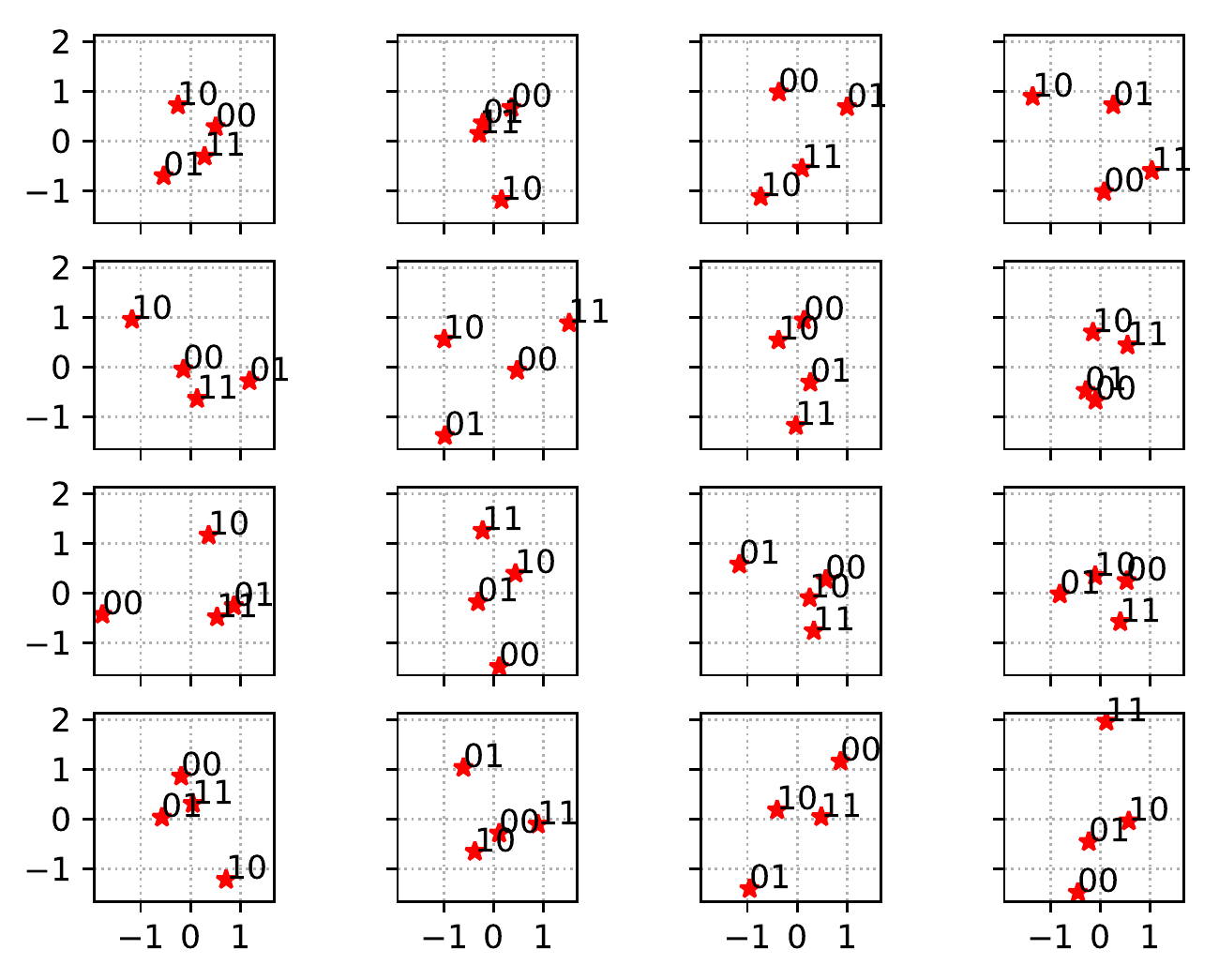}
    \caption{Learned with pML decoder for $k=2$, $L=4$, $(m,n)=(4,4)$}
\end{subfigure}
\caption{Examples of signal constellations learned with: (a) NN decoder, (b) pML decoder. Complex signal depicted for each antenna (across rows) and time slot (across columns).}
\label{fig:codes}
\end{figure*}

The block error rate (BLER) performance results are shown in Figure~\ref{fig:BLER-L2} for $L=2$ and Figure~\ref{fig:BLER-L4} for $L=4$, with the NN decoder results appearing on the left, and the pML decoder results appearing on the right.
Note that for several operating points (seven for $L=2$ and two for $L=4$), the pML results exhibit large error floors, while the NN results generally do not.
At other operating points, the results between NN and pML are similar (although sometimes slightly better or worse).
Due to time constraints, we searched over six times fewer hyperparameters (optimization instances) for the pML decoder experiments, which we believe plays a significant role in the optimization failing in some cases.
For the NN experiments, there were similar optimization failures for other hyperparameters.
Interestingly, despite the orthonormal loss-term, only one operating point ($k=2$, $L=4$, $m=n=2$) resulted in the codebook for the pML decoder converging to orthonormal codewords.
However, we did find that the presence of the of the orthonormal loss-term improved that optimization success rate.
Two examples of learned signal constellations are shown in Figure~\ref{fig:codes}.

\subsection{Neural Network Architectures} \label{sec:archs}

We use two well-known neural network architectures, the multilayer perceptron (MLP) and the Residual MLP (ResMLP)~\cite{he2016deep, huang2017learning}, to realize the neural network-based decoders discussed in Section~\ref{sec:decoders}.

In the MLP architecture, the input vector $\x_0$ is mapped to the output vector $\x_{l+1}$ by applying a series of affine transformations and element-wise, nonlinear operations.
The $l$ hidden (intermediate) layers and output layer (vector) of the network are given by
\begin{align*}
\x_{i+1} := \phi_i(\W_i \x_i + \bias_i), \quad \text{for}\ i \in \{0, \ldots, l\},
\end{align*}
where $\{\W_i, \bias_i\}_{i=0}^l$ are the affine transformation parameters that define the network, and $\phi_i(\cdot)$ denotes the element-wise application the activation function $\phi_i$.
For all of our MLP networks, we used the rectified linear unit (ReLU) for the hidden layers (i.e., $\phi_i(x) := \max(x, 0)$, for $i \in \{0, \ldots, l-1\}$) and the identity function for the output layer (i.e., $\phi_l(x) = x$).
Note that the dimensions of the weight matrices $\W_i$ and bias vectors are constrained by the desired input, output, and hidden layer dimensions.

In the ResMLP architecture, the input vector $\x$ is first mapped to an initial hidden vector $\h_0$ via an affine transformation, i.e.,
\begin{align*}
\h_0 := \W_0 \x + \bias_0.
\end{align*}
Then, over $l$ blocks, the hidden vector is updated according to
\begin{align*}
\h_i := F_{2i}(F_{2i-1}(\h_{i-1})) + \h_{i-1}, \quad \text{for}\ i \in \{1, \ldots, l\},
\end{align*}
where $F_i(\cdot)$ denotes the sequential application of batch-normalization~\cite{ioffe2015batch}, an activation function, and affine transform, as given by
\begin{align*}
F_i(\h) := \W_i \phi_i(\mathrm{BatchNorm}(\h)) + \bias_i.
\end{align*}
Finally, the output is computed as
\[
\y := \W_{2i+1} \phi_{2i+1}(\h_l) + \bias_{2i+1}.
\]

\subsection{Training Procedures} \label{sec:hyper}

We perform the optimization of the objectives given in Section~\ref{sec:obj} with stochastic gradient descent (SGD), specifically the popular Adam~\cite{kingma2014adam} variant, which adaptively adjusts learning rates based on moment estimates.
For each iteration, the expectations are approximated by the empirical mean over a batch of $10,000$ uniformly sampled messages, randomly drawn along with random channel matrices and noise for the transmission of each message.
Training was performed for up to $50,000$ iterations, with early stopping applied to halt training when the objective fails to improve, while saving the best snapshot in terms of BLER.
We implemented these experiments using the Chainer deep learning framework~\cite{chainer2015}.

For the NN decoder, we tried both the MLP and ResMLP architectures across the combination of $l \in \{1, 2, 3\}$ layers/blocks and $\{256, 500, 1000\}$ hidden layer dimensions.
For the pML decoder, the main hyperparameter is just the weight $\lambda$ in the objective function given by~\eqref{eqn:ML-objective}, which we varied across $\lambda \in \{1.0, 3.0, 10.0\}$.
For both decoders, an additional hyperparameter is the SNR used during training simulations, which we non-exhaustively varied from 10~dB to 30~dB in 5~db increments, with one to three SNRs tried for each operating point.


\section{Discussion and Ongoing Work}

Our experiments reevaluated the role of neural networks in learning-based approaches to communications.
We demonstrated that neural networks can be avoided altogether while still realizing the fundamental concept of simulation-driven design optimization.
We also used this approach to show the feasibility of non-coherent MIMO for coherence windows that are as short as two time slots.

Further experimentation and hyperparameter tuning is necessary and part of ongoing work, in order to further confirm our experimental observations.
In particular, we believe that the convergence stability (and possibly the performance) of the pML decoder could be further improved with more tuning, since due to time constraints we had to explore a much smaller hyperparameter space.

The generalized log-likelihood ratio test (GLRT) decoder given by~\cite{koike2010high} does not require the codewords to be orthonormal, which would obviate the need for an orthonormal loss term.
Due to the somewhat increased implementation and computational complexity, applying this GLRT deocder remains ongoing work.


\bibliographystyle{IEEEtran}
\bibliography{ref}

\end{document}